\documentclass[aps,prl,twocolumn,groupedaddress]{revtex4-1}
\usepackage{graphicx}

\begin{document}


\title{Cosmic curvature from de Sitter equilibrium cosmology}


\author{Andreas Albrecht}
\affiliation{The University of Chicago department of Astronomy and
  Astrophysics and\\
Kavli Institute for Cosmological Physics;
5640 South Ellis Avenue; 
Chicago, IL 60637\\
(Sep 2010--May 2011)\\}
\affiliation{University of California at Davis;
Department Of Physics\\
One Shields Avenue;
Davis, CA 95616\\
(Permanent address)}



\begin{abstract}
I show that the de Sitter Equilibrium cosmology generically predicts
observable levels of curvature in the Universe today. The predicted
value of the curvature, $\Omega_k$, depends only on the ratio of the density of
non-relativistic matter to cosmological constant density,
$\rho^0_m/\rho_\Lambda$, and the value of the curvature from the initial
bubble that starts the inflation, $\Omega_k^B$.  The
result is independent of the scale of inflation, the shape of the
potential during inflation, and many other details of the cosmology.
Future cosmological measurements of 
$\rho^0_m/\rho_\Lambda$ and $\Omega_k$ will open up a window on the
very beginning of our Universe and offer an opportunity to support or
falsify the de Sitter Equilibrium cosmology.

\end{abstract}


\maketitle

The de Sitter Equilibrium (``dSE'') cosmology is a framework for
cosmology that pictures the Universe eternally fluctuating in an
equilibrium state.  In this picture phenomena similar to the cosmos we
observe around us come about as fluctuations.  The dSE framework assumes
that the observed cosmic acceleration is driven by a true cosmological
constant $\Lambda$ which causes the Universe to approach a ``de Sitter
space'' at late times when the cosmological constant dominates the
cosmic evolution.  The de Sitter space is the equilibrium state.
It has an entropy given by  $S_\Lambda = \pi (cH_\Lambda^{-1}/l_P)^2$ which is known to be
maximal  \cite{Gibbons:1977mu}, a temperature given by $T_\Lambda = k_B\hbar H_\Lambda$
where $H_\Lambda = 8\pi G/3 \rho_\Lambda \equiv \Lambda/3$ 
is the Hubble constant during the $\Lambda$ dominated phase and $l_P$ is
the Planck length. 
Background on dSE cosmology, including how it evades the notorious
``Boltzmann Brain'' problem of equilibrium cosmologies may be found in  \cite{Albrecht:2004ke,Albrecht:2009vr}.

Cosmic inflation gives an established account of the
very early history of the Universe.  Inflation assumes that the 
early Universe was dominated by the potential energy of a scalar
field, the ``inflaton'', which caused a period of accelerated
cosmic expansion or ``inflation'' before decaying into ordinary matter
through a process called reheating. A simple account of this
inflationary epoch leads to a detailed set of predictions which so far
have been born out by observations  \cite{Komatsu:2010fb}.  However,
to understand inflation fully and make the predictions robust one must put
cosmic inflation into a larger context that accounts for how inflation
starts and assigns relative probabilities to different possible starts to
inflation as well as other starts to the observed Universe
that may not even involve inflation. dSE cosmology gives one way to do
this. 

\begin{figure}
\includegraphics[height=2.6323in,width=3.2in]{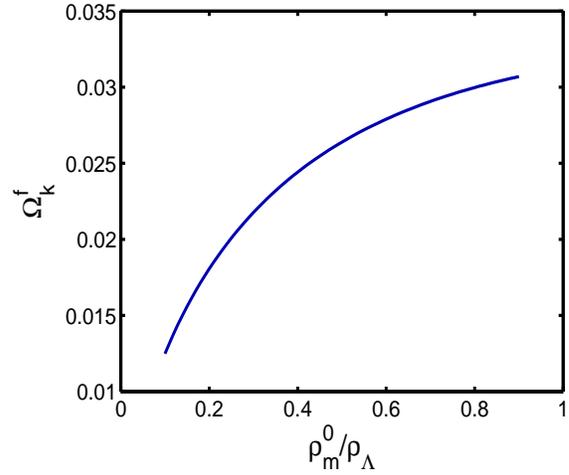}
\caption{\label{fig.result} The  predicted value of $\Omega_k^f$
vs. $\rho_m/\rho_\Lambda$ using the fiducial value $\Omega_k^B = 0.5$.
Predications from other values of the bubble   curvature are given by
$\Omega_k = \Omega_k^f\times(\Omega_k^B/0.5)$  } 
\end{figure}

All ideas for complete cosmological frameworks (including dSE and the
popular ``eternal inflation'' picture) involve some ad hoc
assumptions about how the underlying fundamental physics actually
works \cite{Albrecht:2009vr}. Until we understand which assumptions about the fundamental
physics are correct, the best any of these pictures can hope
to provide is an opportunity for observational tests of one
set of assumptions or another. This paper reports a
prediction of the value of the cosmic curvature, $\Omega_k$, from the dSE picture.
The predicted value depends only on the ratio of the non-relativistic
matter density today $\rho_m^0$ to $\rho_\Lambda$ and is proportional to the
initial curvature ($\Omega_k^B$) provided by
the bubble  that started the period of cosmic inflation.  The
prediction is depicted in Fig.~\ref{fig.result}.

The prediction is interesting for a couple of reasons.  Firstly, the
result is only just consistent with current data
\cite{Komatsu:2010fb}, and uncertainties in 
$\Omega_k$, $ \rho_m$ and $\rho_\Lambda$ will reduce substantially in
the foreseeable future \cite{Albrecht:2006um,*Knox:2006ux}.  Future
data  showing a positive value for $\Omega_k$ would offer strong support
for the dSE picture.  Data consistent with $\Omega_k=0$ with very
small uncertainties (the cosmic variance limit of $\Delta\Omega_k
\approx 0.00001$ may be achievable) would rule out the dSE picture except
for extremely small values of $\Omega_k^B$. Further study of the
initial bubbles could even completely rule out the small $\Omega_k^B$
dSE case, leading to the possibility of fully falsifying the dSE picture.

Secondly, the dSE prediction is interesting for its lack of dependence
on many details of the cosmology. There is no dependence,
for example, on the shape of the inflaton potential during inflation,
the scale of inflation, the specifics and duration of the reheating
and many other factors.  As future data further constrain
cosmological parameters, only $\Omega_k^B$ will be in play, and this
quantity would be subject to the sort of pressures just discussed.

\begin{figure}
\includegraphics[height=2.6323in,width=3.2in]{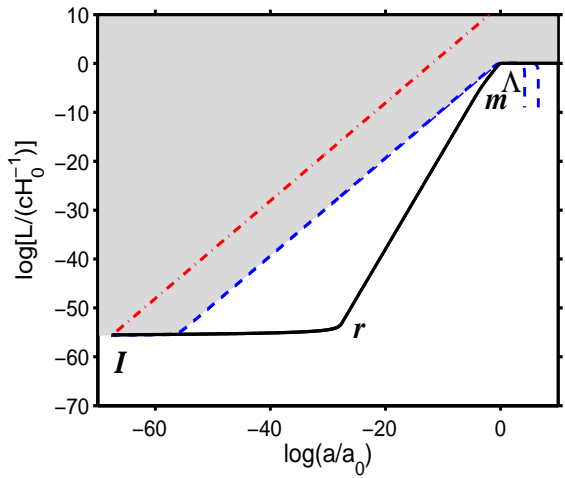}
\caption{\label{fig.Hlength} {\it Solid}: The Hubble length vs. cosmic scale factor
  $a$ (scaled by today's values, $cH_0^{-1}$ and $a_0$
  respectively). The letters mark the times when the density of the inflaton
  ({\bf \it I}),
  relativistic matter ({\bf \it r}), non-relativistic matter ({\bf
    \it m})
  and ({\bf $\Lambda$}) in turn start to dominate Eqn.~\ref{Feq}. 
{\it Dashed}: The
  past horizon of observations deep in the de Sitter era.  The shaded
  region shows events that will never be seen by the observer no
  matter how late the observation is made.  {\it Dot-Dashed}: Maximum
  length scale affected by inflation.} 
\end{figure}

I now derive the result shown in Fig.~\ref{fig.result}.
The Friedmann equation 
\begin{equation}
{H^2} \equiv {\left( {\frac{{\dot a}}{a}} \right)^2} = \frac{{8\pi
    {G}}}{3}\left( {{\rho _I} + {\rho _r} + {\rho _m} + {\rho_k} +
  {\rho_\Lambda }} \right) 
\label{Feq}
\end{equation}
relates the expansion rate $H$ to the (effective) energy densities of,
in order of appearance,  
the inflaton, relativistic matter, 
non-relativistic matter, curvature, and cosmological constant in a
homogeneous and 
isotropic Universe. The scale factor $a$ tracks the cosmic
expansion. For the solutions I consider $a$ is monotonic in time and I
use it as a time variable in what follows. The ``Hubble length''
($\equiv cH^{-1}$) is shown by the solid curve in
Fig.~\ref{fig.Hlength} for the entire history of the Universe in a
standard cosmological picture.  

The dashed curve in Fig.~\ref{fig.Hlength} is the ``past horizon''
\begin{equation}
{h_p}\left( {{a_1}} \right) \equiv {a_1}\int_{{a_1}}^{{a_\Lambda }}
{\frac{{da}}{{{a^2}H}}}    
\label{eqn.hp}
\end{equation}
of
observations at a time deep in the de Sitter era given by $a_\Lambda$.
Specifically, $h_p(a_1)$  is the physical distance at time $a_1$
 between an observer at rest with the expansion and a photon
that will just reach the observer when
$a=a_\Lambda$. Curves are shown for two values of  
$a_\Lambda$ (identifiable by the value of $a$ where each curve drops
toward zero).  The two overlap except right near the respective 
values of $a_\Lambda$.  This is due to the 
event horizon (of size $=cH^{-1}$) that forms in the de Sitter era.
From right to left, the two past horizon curves run up
against the event horizon and then ``exit'' out into the pre-de Sitter
era together.  The past horizon 
for any event deep in the de Sitter era will will do the same. 
The shaded region in Fig.~\ref{fig.Hlength} represents events that will
never be seen by the observer even after waiting an infinitely long
time. 

Even though $H$ appears in the integral defining the past horizon, the
curve takes a simple form $h_p \propto a$ for much of the history of
the Universe, independently of the (possibly complicated) behavior
of $H(a)$.
This is because the evolution of $h_p$ is dominated by the
cosmic expansion for ${h_p} \gg {cH^{ - 1}}$ (in this regime the
cosmic expansion increases the distance $h_p$ at a rate much faster than
$c$). This simple behavior for $h_p(a)$, regardless of the behavior of
$H(a)$ over much of the cosmic history, is central to how the
prediction for $\Omega_k$ remains independent of many details of the
cosmology \footnote{
As defined $h_p(a_i)$ is quite different from
$H^{-1}$ or the past horizon of events {\em during} inflation both of which
are roughly constant functions of $a$ in the inflation era.}.

In the dSE framework the equilibrium state has finite entropy
$S_\Lambda$, and it has been argued that finite $S_\Lambda$
implies the full physical system is finite, describable in a finite
Hilbert space with dimension $e^{S_\Lambda}$~
\cite{Banks:2007ei,*Dyson:2002pf}.  For such 
a finite system, any 
field theoretic description is necessarily approximate and will only
have a finite domain of validity (otherwise an infinite Hilbert space
would be needed). Limitations on the validity of field theory
will necessarily limit scalar field inflation. (When allowed an
unlimited domain of 
inflaton validity, inflation typically leads to ``eternal inflation''
\cite{Linde:1986fd,*Guth:2007ng} which
lasts forever, creates infinite entropy and volume, and incurs problematic
measure issues as a result). The dSE picture reconciles inflation
with the finite entropy by only allowing an amount of inflation
sufficient to fill the horizon of the observer.  This bound prevents
the cosmology from producing more entropy than the maximum value $S_\Lambda$. 
Also, the dSE bound generically keeps inflation far from the ``self
reproduction'' regime that leads to eternal inflation.

The dSE bound originates with the finiteness suggested by the de
Sitter horizon from a truly
constant $\Lambda$ and in that sense is ``holographic'' \cite{Albrecht:2009vr}.
Holographic bounds which do not incorporate $\Lambda$ from today's
acceleration (critical to my argument of finiteness) give much less restrictive
results \cite{Albrecht:2002xs,*ArkaniHamed:2007ky,*Dubovsky:2008rf}.
Interestingly, when re-expressed as a constraint on inflationary
e-foldings the dSE bound looks similar to the bound found in
\cite{Banks:2003pt} which does include $\Lambda$ (although the methods in
\cite{Banks:2003pt} appear different) \cite{PhillipsZZ}.   

By staying strictly finite, the dSE
picture does not have the measure problems of the infinite
inflationary scenarios. According to the dSE bound the early part of
inflation adjacent to the shaded region of Fig.~\ref{fig.Hlength} is
excluded by the breakdown of the field theory description.   The
earliest that inflation, and thus the $cH^{-1}(a)$ curve, is allowed to
start is right on the past 
horizon (dashed) curve.  Since other factors 
exponentially favor inflation starting as early as
possible~\cite{Albrecht:2009vr} I take the dSE bound to be saturated in  
what follows: I start all inflation scenarios ``on the past
horizon'', giving $c(H^i)^{-1} = h_p(a^i)$ where $i$ superscripts designate
the start of inflation~\footnote{Subtleties about bubble details and
  the onset of inflation are absorbed into $\Omega_k^B$ and are
  discussed further in \cite{Albrecht:2011aa}}.

\begin{figure}
\includegraphics[height=2.6323in,width=3.2in]{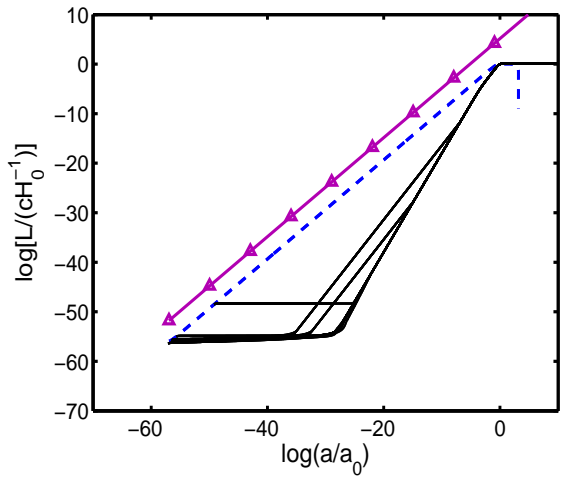}
\caption{\label{fig.MultiInf} {\it Solid}: The Hubble length for many
  different inflation   models, all saturating the dSE bound by
  starting on the past horizon   curve ({\it dashed}).  The 47
  different inflation models (not resolved on this plot)
  represent a wide range of inflation scales, inflaton potentials
  and reheating rates (but use the same values of $\rho_m^0$ and
  $\rho_\Lambda$). Also shown is the   curvature length given by 
  $cH_k^{-1}$ ({\it triangles}). The fact that the past horizon
  (which defines the dSE bound) tracks $cH_k^{-1}$ in such a simple
  manner   over most of the history of the Universe leads to the
  simple   predictions for $\Omega_k$, independently of many details
  of the cosmology. 
}
\end{figure}

Figure~\ref{fig.MultiInf} is similar to Fig.~\ref{fig.Hlength} except that
a multitude of different inflation scenarios are shown on the same
plot.  
There are actually 47 different solid curves (unresolved on the
plot, and discussed in detail in  \cite{Albrecht:2011aa}) which correspond to
changing the inflation potential, the scale of inflation, and the
rate of reheating (in the slowest cases, the reheating only completes
at around $T = 10^{10}K$, just in time for Big Bang Nucleosynthesis).  Each
scenario starts right on the past horizon (dashed line) thus
saturating the past horizon bound.

\begin{figure}
\includegraphics[height=2.6323in,width=3.2in]{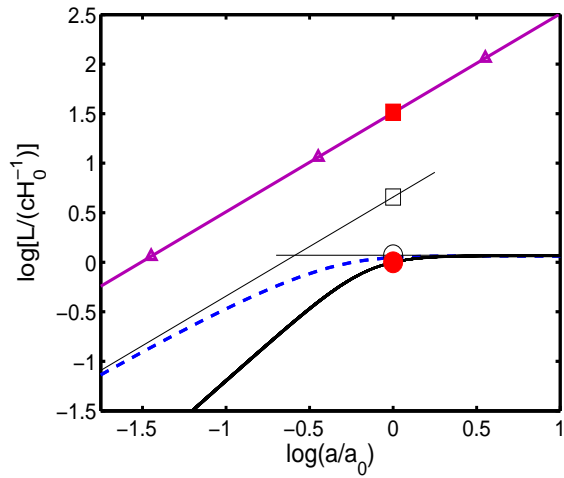}
\caption{\label{fig.MultiInfZoom} This close-up of Fig \ref{fig.MultiInf}
in the vicinity of $a=a_0$ shows the same curves, with additional
curves ({\it thin solid lines}) showing the two asymptotes of the past horizon curve ({\it dashed}).
Markers show quantities evaluated at $a=a_0$ relevant to calculating
$\Omega_k \equiv (H_k/H)^2$
 as discussed in the text. The value of
$cH_k^{-1}$ ({\it filled square}) is related to the {\it open square}
 by $\Omega_k^B$ 
while $H^{-1}$ ({\it filled circle}) is related to the horizontal asymptote
(partly hidden {\it open circle} at $c H^{-1}_\Lambda$) by
Eqn.~\ref{Feq}.  The shape of the past horizon 
curve (Eqn.~\ref{eqn.hp}) quantitatively links the two asymptotes, giving
Eqn.~\ref{eqn.TheAnswer} 
}
\end{figure}

Figure~\ref{fig.MultiInf} also shows the curvature radius $cH^{-1}_k$
($H_k^2 \equiv {8\pi G/3 \rho_k}$), displaying 
information about $\rho_k$ on the plot and helping to illustrate how the
simple dSE prediction comes about.
The initial value $H_k^i$ is
related to the bubble curvature and the initial Hubble parameter $H^i$
by $\Omega_k^B \equiv (H_k^i/H^i)^2$.
Because by definition $\rho_k\propto 1/a^2$, $H_k^{-1} \propto a$, so $cH_k^{-1}$
runs parallel to the past horizon in Fig.~\ref{fig.MultiInf} for most
of the history of the Universe. This parallel behavior is crucial to
my result.  According to the dSE picture, the curve $cH^{-1}$ must start
on the past horizon curve.  The $cH^{-1}_k$ and $h_p$ curves evolve linearly together
until near the current epoch (shown in
Fig.~\ref{fig.MultiInfZoom}), allowing a simple relationship to be
established between $\Omega_k$ and $\Omega_k^B$. 

Figure~\ref{fig.MultiInfZoom} illustrates how the crucial ingredients
needed to compute 
$\Omega_k \equiv (H_k/H)^2$
are all contained in the curves
and asymptotic behaviors discussed above. To the extent that we know
$\rho_m^0$, $\rho_k$ and $\rho_\Lambda$ we know the shape of $H(a)$
around today, since 
$\rho_m$ and $\rho_\Lambda$ (and to a much lesser extent $\rho_k$)
completely dominant Eqn.~\ref{Feq} 
during the current era~\footnote{Evolving dark energy
  could change the shape of $h_p(a)$ but a constant $\Lambda$ is
  essential to the dSE idea.
}.  The quantity $H(a)$ appears in $\Omega_k$ as
well the expression for $h_p(a)$ (Eqn.~\ref{eqn.hp}).  Since $h_p(a)$
only deviates 
from its asymptotic values around the current era, only the 
$\rho$'s listed here  are needed to determine this curve as well. The
quantitative expressions for these various ingredients (all given
above) can be combined to produce the main result:
\begin{equation}
{\Omega _k} = \frac{1}{{{g^2}}}\frac{{\Omega _k^B}}{{\left( {\frac{{\rho _m^0}}{{{\rho _\Lambda }}} + \frac{{\rho _k^0}}{{{\rho _\Lambda }}} + 1} \right)}}
\label{eqn.TheAnswer}
\end{equation}
where 
\begin{equation}
g \left( \frac{\rho _m^0}{\rho _\Lambda }
,\frac{\rho_k^0}{\rho _\Lambda } \right) 
\equiv 
\int_0^\infty  {\frac{{dx}}{{{x^2}\sqrt {{x^{ - 3}}\frac{{\rho
            _m^0}}{{{\rho _\Lambda }}} + {x^{ - 2}}\frac{{\rho
            _k^0}}{{{\rho _\Lambda }}}+1} }}}  
\label{eqn.gDef}
\end{equation}
Due to the appearance of $\rho_k^0$ on the right hand side,
Eqn.~\ref{eqn.TheAnswer} give an implicit equation for 
$\Omega_k\left(\rho_m^0/\rho_\Lambda,\Omega_k^B\right)$ which can
easily be solved numerically to give Fig.~\ref{fig.result}.

Attempts such as dSE
to construct a complete theoretical framework for cosmology are in
a primitive state.   There are a number of ad hoc assumptions that go
into the dSE framework (spelled out in \cite{Albrecht:2009vr}).  The
prediction I report here should be understood in that context.  
Probably
the most popular cosmological framework is eternal inflation. That 
picture has its own particular assumptions, including the validity of the
semiclassical inflaton field theory coupled to Einstein gravity over
an infinite time and infinite volume.  These infinities are critical
to the mechanisms believed to cause eternal inflation to
dominate the cosmos and any breakdown of
these assumptions (such as replacing either of these infinities with
arbitrarily large but finite values) would undermine much of the
current thinking on this subject.  
The measure problem of eternal inflation
that has so far undermined the ability of eternal inflation to
actually make predictions is
related to these infinities, but many are hopeful that this
problem will eventually find a resolution without removing the
infinities that are considered critical to the overall picture
\cite{Linde:2010xz,*Freivogel:2009rf,*Guth:2007ng,*DeSimone:2008if}). 

The dSE framework is a finite alternative to eternal inflation.
The finiteness has its own intrinsic appeal (for
example, dSE replaces assumptions about initial conditions with an equilibrium
state for the Universe), and the finiteness prevents measures from being a
problem. The dSE picture is based on the idea that physics operates
in such a way that the physical world, at its most fundamental, is
described by a finite Hilbert space of dimension
$e^{S_\Lambda}$. One then has to view any field theoretic degrees of
freedom such as the inflaton or those of Einstein gravity as approximate, since
it would take an infinite Hilbert space to describe them fully. The
dSE framework makes assumptions about when the field
theoretic description is a good one and when and how it breaks down.
These assumptions are chosen to give a workable cosmology.  One can
think of the dSE cosmology as an attempt to construct a realistic
finite cosmology by exploiting uncertainties about the underlying
fundamental physics. 

The existence of such great uncertainties may not be satisfying, but it
is the state of the art.  Under these conditions one can hope that
by demanding a realistic cosmology insights might be gained into the
nature of the underlying physics. This project was conceived in this
spirit and it is in this context that I find the result very
interesting.  Unlike other models that give nonzero
$\Omega_k$
\cite{Garriga:1996pg,*Linde:1999wv,*Barnard:2004qm,*Freivogel:2005vv}), 
this result is independent of the shape and scale of the inflaton
potential during inflation, the nature of reheating and many 
other details.  

If future data reveal positive values of $\Omega_k$ close to the
current bounds, that could be seen as support for the dSE picture. 
Such results could be interpreted as constraining the value of 
$\Omega_k^B$, giving a direct window on the tunneling event that created the
Universe we observe.  Further work is needed to understand how low a
value of $\Omega_k^B$ can be tolerated in this picture, but it seems
unlikely that values much smaller than the current bound will make
sense.  If this claim is born out, the result presented here
offers an opportunity to falsify the dSE picture. 

I thank L. Knox, E. Kolb and D. Phillips for
helpful discussions, and the KICP, the 
University of Chicago department of Astronomy and Astrophysics and
NORDITA for hospitality during my sabbatical. This work was supported
in part by  DOE Grant DE-FG03-91ER40674.

\bibliography{dSE}

\begin{thebibliography}{26}%
\makeatletter
\providecommand \@ifxundefined [1]{%
 \@ifx{#1\undefined}
}%
\providecommand \@ifnum [1]{%
 \ifnum #1\expandafter \@firstoftwo
 \else \expandafter \@secondoftwo
 \fi
}%
\providecommand \@ifx [1]{%
 \ifx #1\expandafter \@firstoftwo
 \else \expandafter \@secondoftwo
 \fi
}%
\providecommand \natexlab [1]{#1}%
\providecommand \enquote  [1]{``#1''}%
\providecommand \bibnamefont  [1]{#1}%
\providecommand \bibfnamefont [1]{#1}%
\providecommand \citenamefont [1]{#1}%
\providecommand \href@noop [0]{\@secondoftwo}%
\providecommand \href [0]{\begingroup \@sanitize@url \@href}%
\providecommand \@href[1]{\@@startlink{#1}\@@href}%
\providecommand \@@href[1]{\endgroup#1\@@endlink}%
\providecommand \@sanitize@url [0]{\catcode `\\12\catcode `\$12\catcode
  `\&12\catcode `\#12\catcode `\^12\catcode `\_12\catcode `\%12\relax}%
\providecommand \@@startlink[1]{}%
\providecommand \@@endlink[0]{}%
\providecommand \url  [0]{\begingroup\@sanitize@url \@url }%
\providecommand \@url [1]{\endgroup\@href {#1}{\urlprefix }}%
\providecommand \urlprefix  [0]{URL }%
\providecommand \Eprint [0]{\href }%
\providecommand \doibase [0]{http://dx.doi.org/}%
\providecommand \selectlanguage [0]{\@gobble}%
\providecommand \bibinfo  [0]{\@secondoftwo}%
\providecommand \bibfield  [0]{\@secondoftwo}%
\providecommand \translation [1]{[#1]}%
\providecommand \BibitemOpen [0]{}%
\providecommand \bibitemStop [0]{}%
\providecommand \bibitemNoStop [0]{.\EOS\space}%
\providecommand \EOS [0]{\spacefactor3000\relax}%
\providecommand \BibitemShut  [1]{\csname bibitem#1\endcsname}%
\let\auto@bib@innerbib\@empty
\bibitem [{\citenamefont {Gibbons}\ and\ \citenamefont
  {Hawking}(1977)}]{Gibbons:1977mu}%
  \BibitemOpen
  \bibfield  {author} {\bibinfo {author} {\bibfnamefont {G.~W.}\ \bibnamefont
  {Gibbons}}\ and\ \bibinfo {author} {\bibfnamefont {S.~W.}\ \bibnamefont
  {Hawking}},\ }\href@noop {} {\bibfield  {journal} {\bibinfo  {journal} {Phys.
  Rev.}\ }\textbf {\bibinfo {volume} {D15}},\ \bibinfo {pages} {2738} (\bibinfo
  {year} {1977})}\BibitemShut {NoStop}%
\bibitem [{\citenamefont {Albrecht}\ and\ \citenamefont
  {Sorbo}(2004)}]{Albrecht:2004ke}%
  \BibitemOpen
  \bibfield  {author} {\bibinfo {author} {\bibfnamefont {A.}~\bibnamefont
  {Albrecht}}\ and\ \bibinfo {author} {\bibfnamefont {L.}~\bibnamefont
  {Sorbo}},\ }\href@noop {} {\bibfield  {journal} {\bibinfo  {journal} {Phys.
  Rev.}\ }\textbf {\bibinfo {volume} {D70}},\ \bibinfo {pages} {063528}
  (\bibinfo {year} {2004})},\ \Eprint {http://arxiv.org/abs/hep-th/0405270}
  {hep-th/0405270} \BibitemShut {NoStop}%
\bibitem [{\citenamefont {Albrecht}(2009)}]{Albrecht:2009vr}%
  \BibitemOpen
  \bibfield  {author} {\bibinfo {author} {\bibfnamefont {A.}~\bibnamefont
  {Albrecht}},\ }\href {\doibase 10.1088/1742-6596/174/1/012006} {\bibfield
  {journal} {\bibinfo  {journal} {J. Phys. Conf. Ser.}\ }\textbf {\bibinfo
  {volume} {174}},\ \bibinfo {pages} {012006} (\bibinfo {year} {2009})},\
  \Eprint {http://arxiv.org/abs/0906.1047} {arXiv:0906.1047 [gr-qc]}
  \BibitemShut {NoStop}%
\bibitem [{\citenamefont {Komatsu}\ \emph {et~al.}(2011)\citenamefont {Komatsu}
  \emph {et~al.}}]{Komatsu:2010fb}%
  \BibitemOpen
  \bibfield  {author} {\bibinfo {author} {\bibfnamefont {E.}~\bibnamefont
  {Komatsu}} \emph {et~al.} (\bibinfo {collaboration} {WMAP}),\ }\href
  {\doibase 10.1088/0067-0049/192/2/18} {\bibfield  {journal} {\bibinfo
  {journal} {Astrophys. J. Suppl.}\ }\textbf {\bibinfo {volume} {192}},\
  \bibinfo {pages} {18} (\bibinfo {year} {2011})},\ \Eprint
  {http://arxiv.org/abs/1001.4538} {arXiv:1001.4538 [astro-ph.CO]} \BibitemShut
  {NoStop}%
\bibitem [{\citenamefont {Albrecht}\ \emph {et~al.}(2006)\citenamefont
  {Albrecht} \emph {et~al.}}]{Albrecht:2006um}%
  \BibitemOpen
  \bibfield  {author} {\bibinfo {author} {\bibfnamefont {A.}~\bibnamefont
  {Albrecht}} \emph {et~al.},\ }\href@noop {} {\  (\bibinfo {year} {2006})},\
  \Eprint {http://arxiv.org/abs/astro-ph/0609591} {astro-ph/0609591}
  \BibitemShut {NoStop}%
\bibitem [{\citenamefont {Knox}\ \emph {et~al.}(2006)\citenamefont {Knox} \emph
  {et~al.}}]{Knox:2006ux}%
  \BibitemOpen
  \bibfield  {author} {\bibinfo {author} {\bibfnamefont {L.}~\bibnamefont
  {Knox}} \emph {et~al.},\ }\href {\doibase 10.1086/508605} {\bibfield
  {journal} {\bibinfo  {journal} {Astrophys. J.}\ }\textbf {\bibinfo {volume}
  {652}},\ \bibinfo {pages} {857} (\bibinfo {year} {2006})},\ \Eprint
  {http://arxiv.org/abs/astro-ph/0605536} {arXiv:astro-ph/0605536} \BibitemShut
  {NoStop}%
\bibitem [{Note1()}]{Note1}%
  \BibitemOpen
  \bibinfo {note} {As defined $h_p(a_i)$ is quite different from $H^{-1}$ or
  the past horizon of events {\protect \em during} inflation both of which are
  roughly constant functions of $a$ in the inflation era.}\BibitemShut {Stop}%
\bibitem [{\citenamefont {Banks}(2007)}]{Banks:2007ei}%
  \BibitemOpen
  \bibfield  {author} {\bibinfo {author} {\bibfnamefont {T.}~\bibnamefont
  {Banks}},\ }\href@noop {} {\  (\bibinfo {year} {2007})},\ \Eprint
  {http://arxiv.org/abs/hep-th/0701146} {arXiv:hep-th/0701146} \BibitemShut
  {NoStop}%
\bibitem [{\citenamefont {Dyson}\ \emph {et~al.}(2002)\citenamefont {Dyson},
  \citenamefont {Kleban},\ and\ \citenamefont {Susskind}}]{Dyson:2002pf}%
  \BibitemOpen
  \bibfield  {author} {\bibinfo {author} {\bibfnamefont {L.}~\bibnamefont
  {Dyson}}, \bibinfo {author} {\bibfnamefont {M.}~\bibnamefont {Kleban}}, \
  and\ \bibinfo {author} {\bibfnamefont {L.}~\bibnamefont {Susskind}},\
  }\href@noop {} {\bibfield  {journal} {\bibinfo  {journal} {JHEP}\ }\textbf
  {\bibinfo {volume} {10}},\ \bibinfo {pages} {011} (\bibinfo {year} {2002})},\
  \Eprint {http://arxiv.org/abs/hep-th/0208013} {hep-th/0208013} \BibitemShut
  {NoStop}%
\bibitem [{\citenamefont {Linde}(1986)}]{Linde:1986fd}%
  \BibitemOpen
  \bibfield  {author} {\bibinfo {author} {\bibfnamefont {A.~D.}\ \bibnamefont
  {Linde}},\ }\href {\doibase 10.1016/0370-2693(86)90611-8} {\bibfield
  {journal} {\bibinfo  {journal} {Phys. Lett.}\ }\textbf {\bibinfo {volume}
  {B175}},\ \bibinfo {pages} {395} (\bibinfo {year} {1986})}\BibitemShut
  {NoStop}%
\bibitem [{\citenamefont {Guth}(2007)}]{Guth:2007ng}%
  \BibitemOpen
  \bibfield  {author} {\bibinfo {author} {\bibfnamefont {A.~H.}\ \bibnamefont
  {Guth}},\ }\href {\doibase 10.1088/1751-8113/40/25/S25} {\bibfield  {journal}
  {\bibinfo  {journal} {J. Phys.}\ }\textbf {\bibinfo {volume} {A40}},\
  \bibinfo {pages} {6811} (\bibinfo {year} {2007})},\ \Eprint
  {http://arxiv.org/abs/hep-th/0702178} {arXiv:hep-th/0702178} \BibitemShut
  {NoStop}%
\bibitem [{\citenamefont {Albrecht}\ \emph {et~al.}(2002)\citenamefont
  {Albrecht}, \citenamefont {Kaloper},\ and\ \citenamefont
  {Song}}]{Albrecht:2002xs}%
  \BibitemOpen
  \bibfield  {author} {\bibinfo {author} {\bibfnamefont {A.~J.}\ \bibnamefont
  {Albrecht}}, \bibinfo {author} {\bibfnamefont {N.}~\bibnamefont {Kaloper}}, \
  and\ \bibinfo {author} {\bibfnamefont {Y.-S.}\ \bibnamefont {Song}},\
  }\href@noop {} {\  (\bibinfo {year} {2002})},\ \Eprint
  {http://arxiv.org/abs/hep-th/0211221} {arXiv:hep-th/0211221} \BibitemShut
  {NoStop}%
\bibitem [{\citenamefont {Arkani-Hamed}\ \emph {et~al.}(2007)\citenamefont
  {Arkani-Hamed}, \citenamefont {Dubovsky}, \citenamefont {Nicolis},
  \citenamefont {Trincherini},\ and\ \citenamefont
  {Villadoro}}]{ArkaniHamed:2007ky}%
  \BibitemOpen
  \bibfield  {author} {\bibinfo {author} {\bibfnamefont {N.}~\bibnamefont
  {Arkani-Hamed}}, \bibinfo {author} {\bibfnamefont {S.}~\bibnamefont
  {Dubovsky}}, \bibinfo {author} {\bibfnamefont {A.}~\bibnamefont {Nicolis}},
  \bibinfo {author} {\bibfnamefont {E.}~\bibnamefont {Trincherini}}, \ and\
  \bibinfo {author} {\bibfnamefont {G.}~\bibnamefont {Villadoro}},\ }\href@noop
  {} {\bibfield  {journal} {\bibinfo  {journal} {JHEP}\ }\textbf {\bibinfo
  {volume} {05}},\ \bibinfo {pages} {055} (\bibinfo {year} {2007})},\ \Eprint
  {http://arxiv.org/abs/0704.1814} {arXiv:0704.1814 [hep-th]} \BibitemShut
  {NoStop}%
\bibitem [{\citenamefont {Dubovsky}\ \emph {et~al.}(2009)\citenamefont
  {Dubovsky}, \citenamefont {Senatore},\ and\ \citenamefont
  {Villadoro}}]{Dubovsky:2008rf}%
  \BibitemOpen
  \bibfield  {author} {\bibinfo {author} {\bibfnamefont {S.}~\bibnamefont
  {Dubovsky}}, \bibinfo {author} {\bibfnamefont {L.}~\bibnamefont {Senatore}},
  \ and\ \bibinfo {author} {\bibfnamefont {G.}~\bibnamefont {Villadoro}},\
  }\href {\doibase 10.1088/1126-6708/2009/04/118} {\bibfield  {journal}
  {\bibinfo  {journal} {JHEP}\ }\textbf {\bibinfo {volume} {04}},\ \bibinfo
  {pages} {118} (\bibinfo {year} {2009})},\ \Eprint
  {http://arxiv.org/abs/0812.2246} {arXiv:0812.2246 [hep-th]} \BibitemShut
  {NoStop}%
\bibitem [{\citenamefont {Banks}\ and\ \citenamefont
  {Fischler}(2003)}]{Banks:2003pt}%
  \BibitemOpen
  \bibfield  {author} {\bibinfo {author} {\bibfnamefont {T.}~\bibnamefont
  {Banks}}\ and\ \bibinfo {author} {\bibfnamefont {W.}~\bibnamefont
  {Fischler}},\ }\href@noop {} {\  (\bibinfo {year} {2003})},\ \Eprint
  {http://arxiv.org/abs/astro-ph/0307459} {arXiv:astro-ph/0307459 [astro-ph]}
  \BibitemShut {NoStop}%
\bibitem [{Phi()}]{PhillipsZZ}%
  \BibitemOpen
  \href@noop {} {}\bibinfo {note} {Daniel Phillips, private Communication
  2011}\BibitemShut {NoStop}%
\bibitem [{Note2()}]{Note2}%
  \BibitemOpen
  \bibinfo {note} {Subtleties about bubble details and the onset of inflation
  are absorbed into $\Omega _k^B$ and are discussed further in \cite
  {Albrecht:2011aa}}\BibitemShut {NoStop}%
\bibitem [{\citenamefont {Albrecht}(2011)}]{Albrecht:2011aa}%
  \BibitemOpen
  \bibfield  {author} {\bibinfo {author} {\bibfnamefont {A.}~\bibnamefont
  {Albrecht}},\ }\href@noop {} {\  (\bibinfo {year} {2011})},\ \bibinfo {note}
  {in Preparation}\BibitemShut {NoStop}%
\bibitem [{Note3()}]{Note3}%
  \BibitemOpen
  \bibinfo {note} {Evolving dark energy could change the shape of $h_p(a)$ but
  a constant $\Lambda $ is essential to the dSE idea.}\BibitemShut {Stop}%
\bibitem [{\citenamefont {Linde}\ and\ \citenamefont
  {Noorbala}(2010)}]{Linde:2010xz}%
  \BibitemOpen
  \bibfield  {author} {\bibinfo {author} {\bibfnamefont {A.}~\bibnamefont
  {Linde}}\ and\ \bibinfo {author} {\bibfnamefont {M.}~\bibnamefont
  {Noorbala}},\ }\href {\doibase 10.1088/1475-7516/2010/09/008} {\bibfield
  {journal} {\bibinfo  {journal} {JCAP}\ }\textbf {\bibinfo {volume} {1009}},\
  \bibinfo {pages} {008} (\bibinfo {year} {2010})},\ \Eprint
  {http://arxiv.org/abs/1006.2170} {arXiv:1006.2170 [hep-th]} \BibitemShut
  {NoStop}%
\bibitem [{\citenamefont {Freivogel}\ and\ \citenamefont
  {Kleban}(2009)}]{Freivogel:2009rf}%
  \BibitemOpen
  \bibfield  {author} {\bibinfo {author} {\bibfnamefont {B.}~\bibnamefont
  {Freivogel}}\ and\ \bibinfo {author} {\bibfnamefont {M.}~\bibnamefont
  {Kleban}},\ }\href {\doibase 10.1088/1126-6708/2009/12/019} {\bibfield
  {journal} {\bibinfo  {journal} {JHEP}\ }\textbf {\bibinfo {volume} {12}},\
  \bibinfo {pages} {019} (\bibinfo {year} {2009})},\ \Eprint
  {http://arxiv.org/abs/0903.2048} {arXiv:0903.2048 [hep-th]} \BibitemShut
  {NoStop}%
\bibitem [{\citenamefont {De~Simone}\ \emph {et~al.}(2010)\citenamefont
  {De~Simone} \emph {et~al.}}]{DeSimone:2008if}%
  \BibitemOpen
  \bibfield  {author} {\bibinfo {author} {\bibfnamefont {A.}~\bibnamefont
  {De~Simone}} \emph {et~al.},\ }\href {\doibase 10.1103/PhysRevD.82.063520}
  {\bibfield  {journal} {\bibinfo  {journal} {Phys. Rev.}\ }\textbf {\bibinfo
  {volume} {D82}},\ \bibinfo {pages} {063520} (\bibinfo {year} {2010})},\
  \Eprint {http://arxiv.org/abs/0808.3778} {arXiv:0808.3778 [hep-th]}
  \BibitemShut {NoStop}%
\bibitem [{\citenamefont {Garriga}(1996)}]{Garriga:1996pg}%
  \BibitemOpen
  \bibfield  {author} {\bibinfo {author} {\bibfnamefont {J.}~\bibnamefont
  {Garriga}},\ }\href {\doibase 10.1103/PhysRevD.54.4764} {\bibfield  {journal}
  {\bibinfo  {journal} {Phys. Rev.}\ }\textbf {\bibinfo {volume} {D54}},\
  \bibinfo {pages} {4764} (\bibinfo {year} {1996})},\ \Eprint
  {http://arxiv.org/abs/gr-qc/9602025} {arXiv:gr-qc/9602025} \BibitemShut
  {NoStop}%
\bibitem [{\citenamefont {Linde}\ \emph {et~al.}(1999)\citenamefont {Linde},
  \citenamefont {Sasaki},\ and\ \citenamefont {Tanaka}}]{Linde:1999wv}%
  \BibitemOpen
  \bibfield  {author} {\bibinfo {author} {\bibfnamefont {A.~D.}\ \bibnamefont
  {Linde}}, \bibinfo {author} {\bibfnamefont {M.}~\bibnamefont {Sasaki}}, \
  and\ \bibinfo {author} {\bibfnamefont {T.}~\bibnamefont {Tanaka}},\ }\href
  {\doibase 10.1103/PhysRevD.59.123522} {\bibfield  {journal} {\bibinfo
  {journal} {Phys. Rev.}\ }\textbf {\bibinfo {volume} {D59}},\ \bibinfo {pages}
  {123522} (\bibinfo {year} {1999})},\ \Eprint
  {http://arxiv.org/abs/astro-ph/9901135} {arXiv:astro-ph/9901135} \BibitemShut
  {NoStop}%
\bibitem [{\citenamefont {Barnard}\ and\ \citenamefont
  {Albrecht}(2004)}]{Barnard:2004qm}%
  \BibitemOpen
  \bibfield  {author} {\bibinfo {author} {\bibfnamefont {M.}~\bibnamefont
  {Barnard}}\ and\ \bibinfo {author} {\bibfnamefont {A.}~\bibnamefont
  {Albrecht}},\ }\href@noop {} {\  (\bibinfo {year} {2004})},\ \Eprint
  {http://arxiv.org/abs/hep-th/0409082} {arXiv:hep-th/0409082} \BibitemShut
  {NoStop}%
\bibitem [{\citenamefont {Freivogel}\ \emph {et~al.}(2006)\citenamefont
  {Freivogel} \emph {et~al.}}]{Freivogel:2005vv}%
  \BibitemOpen
  \bibfield  {author} {\bibinfo {author} {\bibfnamefont {B.}~\bibnamefont
  {Freivogel}} \emph {et~al.},\ }\href {\doibase 10.1088/1126-6708/2006/03/039}
  {\bibfield  {journal} {\bibinfo  {journal} {JHEP}\ }\textbf {\bibinfo
  {volume} {03}},\ \bibinfo {pages} {039} (\bibinfo {year} {2006})},\ \Eprint
  {http://arxiv.org/abs/hep-th/0505232} {arXiv:hep-th/0505232} \BibitemShut
  {NoStop}%
\end{thebibliography}%

\end{document}